\documentclass[a4paper]{article}
\usepackage[margin=25mm]{geometry}
\usepackage{amsfonts}
\usepackage{amssymb}
\usepackage{graphicx}
\usepackage{svg}
\usepackage{amsmath}
\usepackage{caption}
\usepackage[labelfont={small,bf}]{caption}
\usepackage[font={small}]{caption}
\usepackage{subcaption}
\usepackage{hyperref}
\title{Unraveling biochemical spatial patterns: machine learning approaches to the inverse problem of Turing patterns}
\author{{Antonio Matas-Gil}$^{1}$, {Robert G. Endres}$^{1\dagger}$}

\date{\small $^{1}$Department of Life Sciences \& Centre for Integrative Systems Biology and Bioinformatics, Imperial College London, London SW7 2BU, United Kingdom\\\vspace{0.3cm}$^\dagger$r.endres@imperial.ac.uk}   

\begin{document}
\maketitle

\begin{abstract}

The diffusion-driven Turing instability is a potential mechanism for spatial pattern formation in numerous biological and chemical systems. However, engineering these patterns and demonstrating that they are produced by this mechanism is challenging. To address this, we aim to solve the inverse problem in artificial and experimental Turing patterns. This task is challenging since high levels of noise corrupt the patterns and slight changes in initial conditions can lead to different patterns. We used both least squares to explore the problem and physics-informed neural networks to build a noise-robust method. We elucidate the functionality of our network in scenarios mimicking biological noise levels and showcase its application through a prototype involving an experimentally obtained chemical pattern. The findings reveal the significant promise of machine learning in steering the creation of synthetic patterns in bioengineering, thereby advancing our grasp of morphological intricacies within biological systems while acknowledging existing limitations.

\end{abstract}

\section*{Introduction}

Spatial patterns are prevalent in biological systems, including gene expression in microbial communities, developing embryos, as well as skin and fur patterns in adult animals. A leading mechanism for pattern formation is the diffusion-driven instability in reaction-diffusion models, as proposed by Alan Turing \cite{Turing1952} and extended by others \cite{Schnakenberg1979, giererMeinhardt, brusselator, FitzHugh1961, Nagumo1962}. These models typically describe interacting and diffusing activator and inhibitor chemical species through sets of coupled partial-differential equations (PDEs), although simulation methods are also available. While Turing patterns were originally observed in chemical systems \cite{chemTP1990,ChemPat2019} it has been challenging to conclusively demonstrate this mechanism in developmental systems such as digit formation \cite{digitPatterning}, zebra-fish skin pigment patterning \cite{fishPatterning} and hair spacing in mice \cite{hairPatterning} (for recent reviews see \cite{review1,review2}). Additionally, building Turing patterns from bottom-up with synthetic circuits has proven difficult  \cite{stochasticTuring, Sekine2018}. Identified issues are that the Turing mechanism is exceedingly simple compared to biological regulatory pathways, highly sensitive to changes in model parameters \cite{SCHOLES2019243}, and experimental control over parameter tuning is limited. Even if patterns can be produced experimentally, the challenge remains of linking them to the corresponding parameters of a candidate model to support it as an actual Turing pattern rather than an experimental artifact. Further complications arise from the relatively limited amount of data available in developmental systems, which is often corrupted by measurement and imaging noise, as well as the strong pattern variability observed in microbial systems. Addressing these issues would greatly benefit from the ability to robustly estimate parameters from given patterns and candidate models.

To address the issues that arise in an experimental setting, the inverse problem can be initially approached using artificial data from numerical simulations instead of actual experimental data. This involves generating random initial conditions and evolving the reaction-diffusion model in space and time, allowing for a systematic study of the effects of data size and noise. However, solving the inverse problem remains a challenging task even with this simplification, as the same model parameters can produce different patterns, resulting in a many-to-one inversion problem  \cite{Murray2}. This occurs because patterns are highly sensitive to initial conditions, and even slight alterations can cause noticeable changes in the final pattern obtained. These changes do not affect the type of pattern obtained, such as spots or stripes, but do alter the location and shape of the pattern elements. As a result, previous research has ruled out direct minimization of the mean-squared displacement between data and model output for parameter fitting, considering it an ill-posed problem  \cite{Kazarnikov2020}. Instead, focus has been on more advanced approaches, including Bayesian inference and other statistical tools \cite{Kazarnikov2020,CampilloBayes} and different machine-learning techniques such as support vector machines, Kernel-based methods, and neural networks \cite{schnorrimperial}, as well as optimal control theory \cite{Garvie2010}. However, despite some success, these approaches suffer from requiring thousands of training images \cite{CampilloBayes}, sensitivity to noise in the patterns \cite{schnorrimperial}, ad hoc cost or loss functions for quantifying the quality of fit \cite{Kazarnikov2020}, or the requirement to fix some parameters and have knowledge of initial conditions and the pattern evolution \cite{Garvie2010}.

As the inverse Turing problem remains largely unresolved, there is a need to develop new robust approaches, particularly when dealing with small and noisy data. Although conceptually straightforward, the least squares (LS) approach has not been extensively explored in this context. This method requires the definition of a parameter-dependent loss function, which can be the residuals of the model equations \cite{hastie01statisticallearning}. In contrast, deep learning methods, such as neural networks, are usually more reliable when it comes to noise, even though they are computationally expensive to train. A key property that makes neural networks a promising tool for approaching this problem is the universal approximation theorem, which states that given enough parameters, neural networks are capable of approximating any continuous function, including spatial patterns \cite{HORNIK1989359,RBFunivApp}. Additionally, physics-informed neural networks (PINNs) incorporate physical constraints, such as the model equations that should hold for the data, thereby helping to regularize the training \cite{KarniadakisPINNs} and making PINNs more intuitive than black-box neural networks. An added benefit of PINNs is that they do not require large amounts of data; in one case, a single simulation output was sufficient to solve the inverse problem  \cite{Cavanagh2021}. Model parameters are regarded as parameters of the neural network, resulting in the computational cost of the whole method being of the same order as the standalone function approximation. This makes PINNs superb candidates for solving the inverse problem in Turing patterns.

Here, we explore basic LS applied to the PDE residuals and advanced PINNs to address the inverse problem - given Turing patterns and candidate models, we aim to recover the model parameters. Using our first approach, we find that LS is computationally inexpensive, but that it requires mostly exact (albeit little) data and hence does not allow significant noise or using a similar pattern produced from a different model. Applying this methodology to small regions of patterns still allows us to recover the parameters, and we identify the minimum number of necessary pixels. Our second approach uses physics-informed neural networks with a radial basis function (RBF) architecture to approximate the patterns, referred to as RBF physics-informed neural networks (RBF-PINNs). This method remedies many of the issues from the first method at a two-order of magnitude higher computational costs, allowing us to obtain accurate results up to a 10-20\% relative noise given a single snapshot of a Turing pattern, even in an experimental setting with chemical Turing patterns. Our least-squares results show a new perspective on the Turing robustness problem, that a pattern can act like a barcode to a specific parameter combination, given the key of the correct model. Our RBF-PINNs are a promising method to solve the inverse problem, and to guide future experiments in bioengineering in the development of synthetic tissues.
\section*{Results}\label{sec:results}
\subsection*{Models}
\begin{figure*}[t]
\includegraphics[width=\textwidth]{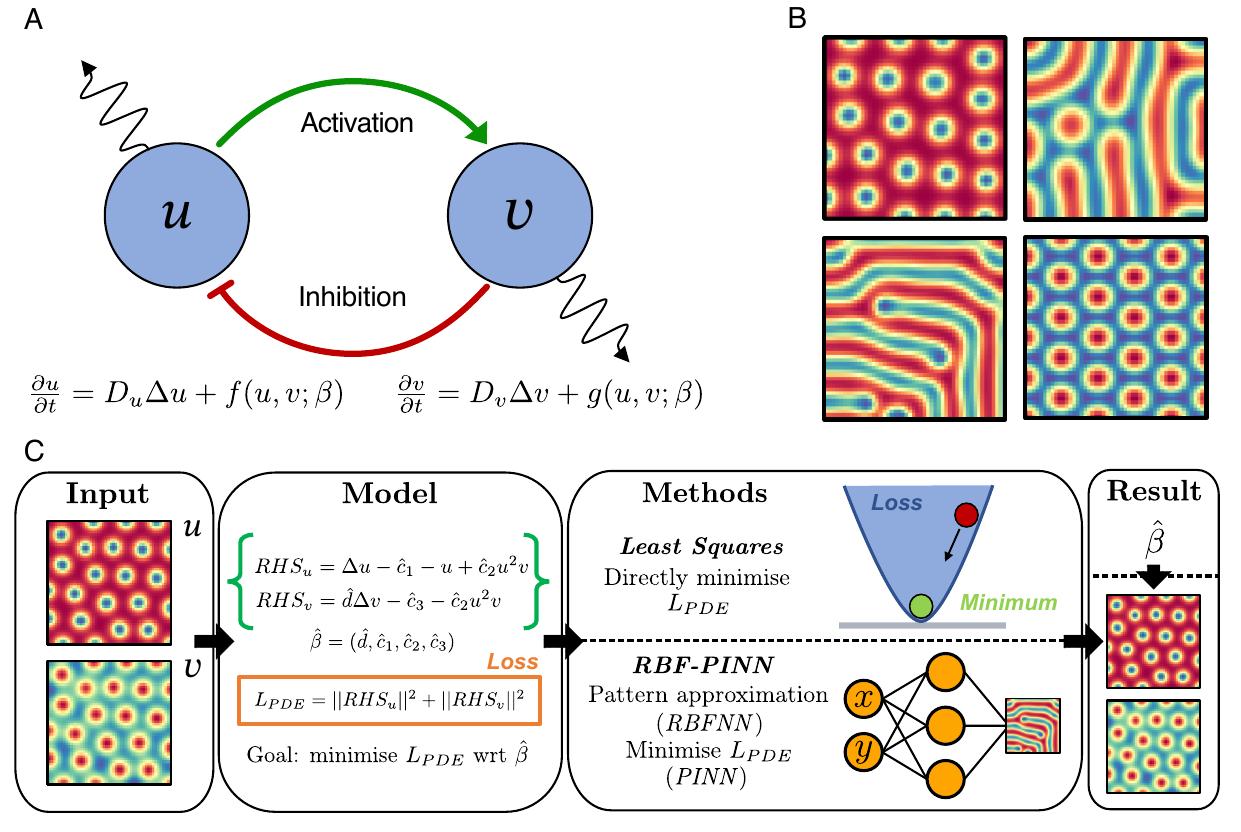}
\caption{\textbf{Turing patterns and methods for the inverse problem.} \textbf{(A)} Network of canonical Turing pattern with a short-range activator and a long-range inhibitor. \textbf{(B)} Turing patterns from different models. The top-left pattern showing spots is produced with the Schnakenberg model, the top-right pattern showing labyrinths with FitzHugh-Nagumo and the bottom row, also showing spots and labyrinths, with the Brusselator and two different parameter sets. \textbf{(C)} Methodology of the paper. Starting from a Turing patterns, we build a PDE loss which is minimized with respect to the parameters $\beta$. This is done using two different methods, LS and RBF-PINNs.}
\label{fig1}
\end{figure*}

Both of our approaches, LS and RBF-PINNs, utilize discrete Turing patterns as inputs to infer the parameters of a candidate model. In this work, the candidate model refers to the true model that was numerically solved to produce the images using the finite-differences algorithm (see \nameref{sec:methods}). We explored several models, all following the idea of activator-inhibitor dynamics shown in Fig. \ref{fig1}A but before delving into specific ones, we present the general two-component reaction-diffusion model for concentrations $u$ and $v$, which depend on space and time, as follows:
\begin{align}\label{general_model}
\begin{split}
    u_t &= D_u \Delta u  + f(u,v) \\
    v_t &= D_v \Delta v + g(u,v)
\end{split}
\end{align}
where $u_t$ and $v_t$ denote partial differentiation with respect to time and the functions $f(u,v)$ and $g(u,v)$ are non-linear reaction kinetics. Depending on the specific  $f$ and $g$ functions, different models can be identified. We will focus on three of these models. First, the Schnakenberg model \cite{Schnakenberg1979}, which has a total of 6 parameters with 4 of them inside the functions $f$ and $g$ and the other two given by the diffusion coefficients:

\begin{align}\label{schnakmodel}
\begin{split}
	f(u,v)&= c_1 - c_2 u + c_3 u^2 v, \\ 
        g(u,v) &= c_4 - c_3 u^2 v
 \end{split}
\end{align}

Second, we have the FitzHugh-Nagumo model \cite{FitzHugh1961, Nagumo1962}, which has 5 parameters: 

\begin{align}\label{fnmodel}
\begin{split}
	f(u,v)&= c_1(v-c_2 u) ,\\
        g(u,v) &= -u+c_3 v - v^3
 \end{split}
\end{align}

Third, the Brusselator model \cite{brusselator}, with a total of 4 parameters:

\begin{align}\label{brusmodel}
\begin{split}
	f(u,v)&= c_1-(c_2+1) u + u^2 v ,\\
        g(u,v) &= c_2 u - u^2 v
\end{split}
\end{align}

We numerically solve these PDE models on a square domain with zero-flux boundary conditions, resulting in patterns similar to Fig. \ref{fig1}B. There are different types of patterns, e.g. dots (Fig. \ref{fig1}B, top left), and labyrinths (Fig. \ref{fig1}B, top right). The type is mostly dependent on the model that we choose: FitzHugh-Nagumo produces labyrinths while Schnakenberg produces dots; but some models, like the Brusselator, can produce several types of patterns depending on the parameters. Parameters for the Brusselator and the FitzHugh-Nagumo models can be found in \cite{Kazarnikov2020}, while those for the Schnakenberg model are in \cite{Maini2012}. Different initial conditions will produce different patterns of the same type, but fixing these will produce the same one. Hence, we can conceptually think of the parameters of the model and the initial conditions as the `variables' that produce a given pattern. Since we are mostly interested in the parameters, we used the same initial conditions for all patterns. This eases comparison between the patterns produced by different methods. As a result, there is a more direct relationship between the parameters and the patterns. The inverse problem consists of inverting this relationship by recovering the parameters that generate a given pattern, as outlined in Fig. \ref{fig1}C.

\subsection*{Non-dimensionalization}

Depending on the model, the inverse problem described above may be unsolvable because it may have many solutions. For example, take the Schnakenberg model, Eq. \ref{general_model} with reaction kinetics given by Eq. \ref{schnakmodel}, and assume $u$ and $v$ are steady-state patterns in time satisfying the PDE (Eq. \ref{general_model}) for a given set of parameters. As a result, both the left-hand side (LHS) and right-hand side (RHS) of the PDE are zero. If we multiply all parameters by a constant $k$ and substitute them back in the equation, the resulting set of parameters is also a solution of the inverse problem; since we can take $k$ as a common factor, we again obtain $RHS = 0$. Furthermore, if we let $k=0$ we arrive at a trivial solution where all parameters are zero. To avoid this, one remedy is to fix some parameters so that there is only a single solution. However, we found that non-dimensionalizing the equations to decrease the number of parameters is preferable, since this does not require us to make any assumptions on the parameters. As an added benefit, this reduces the number of parameters of the model.

We can write the previous models in non-dimensional form as follows. For the Schnakenberg model we obtain:

\begin{equation}
    u_t = \Delta u  + c_1 -  u +c_2 u^2 v  \qquad v_t = d \Delta v + c_3 - c_2 u^2 v,\label{NonDimS}\\
\end{equation}

The FitzHugh-Nagumo model can be rewritten as:
\begin{equation}
    u_t = \Delta u  + v - c_1 u \qquad v_t = d \Delta v + c_2 v -c_3 u -v^3,\label{NonDimFN}\\
\end{equation}

And the non-dimensional form of the Brusselator model is:

\begin{equation}
    u_t = d \Delta u  + 1 -  u + c_1 u^2 v \qquad v_t =  \Delta v  + c_2 u -c_1 u^2 v\label{NonDimBruss}
\end{equation}

where in all equations $d$ is the ratio of the diffusion coefficients and the rest of parameters are non-dimensional. Note that even though we have not changed the notation of the parameters, they are different from the ones in Eqs. \ref{schnakmodel}, \ref{brusmodel} and \ref{fnmodel}. For the definition of the new variables and parameters for the three models see Supplementary Information section 2.

\subsection*{Least squares} 
Our first approach to solve the inverse problem is based on fitting the parameters to the PDE equations, by fixing the concentrations $u$ and $v$ as given by the Turing patterns we aim to reproduce. If we assume the pattern to be a steady state of the PDE in e.g. Eq. \ref{NonDimS}, then $u_t=v_t=0$. We can now consider the RHS of Eq. \ref{NonDimS}. Because we assume we have access to the patterns $u,v$ satisfying the PDE, we can fix them and treat the RHS as a function dependent only on the parameters, which will be zero for the combination of parameters that we want to find. Since our patterns are discrete in space (we can think of them as images with pixels), $\mathbf{u}$ and $\mathbf{v}$ are matrices, so we can write its elements as $u_{ij}$ and $v_{ij}$ for $i,j=1,2,...,N$, where $N$ is the number of rows and columns in the pattern. This makes it possible to write and solve the problem using LS, as we can now formulate the problem as $X\beta=Y$, where $X$ is the design matrix, $\beta$ is the vector of parameters and $Y$ are all terms that are not parameter dependent. (For an example, see the \nameref{sec:methods} section.) This LS minimization is equivalent to minimizing the squared of the Frobenius norm (element-wise $L_2$ norm for matrices) for both equations, which we can write as:

 \begin{equation}
     L(\boldsymbol{\beta}) = ||\beta_1 \Delta \mathbf{u} +f(\boldsymbol{\beta})||_F^2+||\Delta \mathbf{v} +g(\boldsymbol{\beta})||_F^2
 \end{equation}

\begin{figure*}[t]
\includegraphics[width=\textwidth]{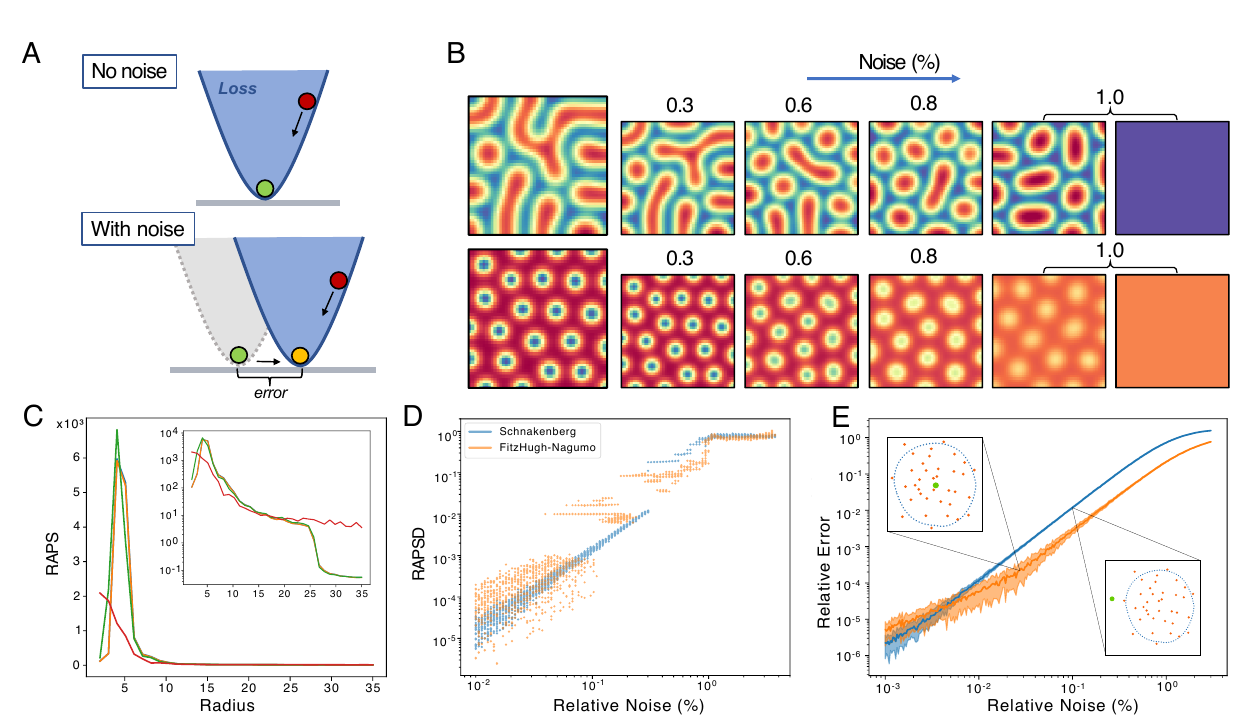}

\caption{\textbf{Least squares for parameter inference.} \textbf{(A)} In LS, noise changes the loss function that we are minimizing and the optimal value shifts, leading to some error in the inferred parameters. \textbf{(B)} Resulting patterns obtained from LS with different noise levels. After corrupting the original pattern with relative noise of different levels, parameters are obtained using LS and the model is solved with the inferred parameters to obtain the new patterns shown. The upper row corresponds to the FitzHugh-Nagumo model, the lower row to the Schnakenberg model. Enlarged patterns on the left are the original ones. It can be observed that at 1\% noise, the predicted patterns are noticeably different from the original ones across the models, and that we can obtain several patterns at the same level of noise. \textbf{(C)} Radially averaged power spectra (RAPS) obtained for different patterns recovered at different noise levels (shown in (B)) from the Schnakenberg model. Red corresponds to 1\% noise, green to 0.6\%, orange to 0.3\% and blue is the original pattern. Inset shows the same plot but the $y$ axis in log-scale. \textbf{(D)} Scatter plot of RAPS differences for different relative noise levels with noticeable sudden increases due to discrete changes in the predicted Turing patterns. \textbf{(E)} Average relative error of inferred parameters for different relative noise levels in the Schnakenberg and FitzHugh-Nagumo models, with a sketch of the parameter space explaining the difference in spread. Green point represents original set of parameters values and orange points are the different optimal sets resulting from the minimization. The bigger standard deviation in relative error occurs when the optimal sets are close to the original so the scale of the error changes drastically (can be very close or far), while the smaller standard deviation occurs when the optimal points are further away and the error is always on a similar scale.}
\label{fig2}
\end{figure*}

The LS method has previously been regarded in the literature \cite{HORNIK1989359}, but to our knowledge it has not been thoroughly investigated. The Laplacian is approximated using a second-order finite difference method, and the no-flux boundary conditions are taken to be the same as used to simulate the pattern. Since we assume the Turing pattern is at steady state, so that $u_t = 0$, we refer to this method as a steady-state approximation. The LS method can also be extended to dynamic patterns, e.g  when having two samples of the time evolution of the system, say $u_1$ and $u_2$ at $t_1$ and $t_2$, respectively. In this case, we can no longer say that $u_t = 0$, but instead approximate this partial derivative using finite differences so that we obtain $u_t \approx \frac{u_2-u_1}{t_2-t_1}$ (see \nameref{sec:discussion}).

\begin{figure*}[t]
\includegraphics[width=0.75\textwidth]{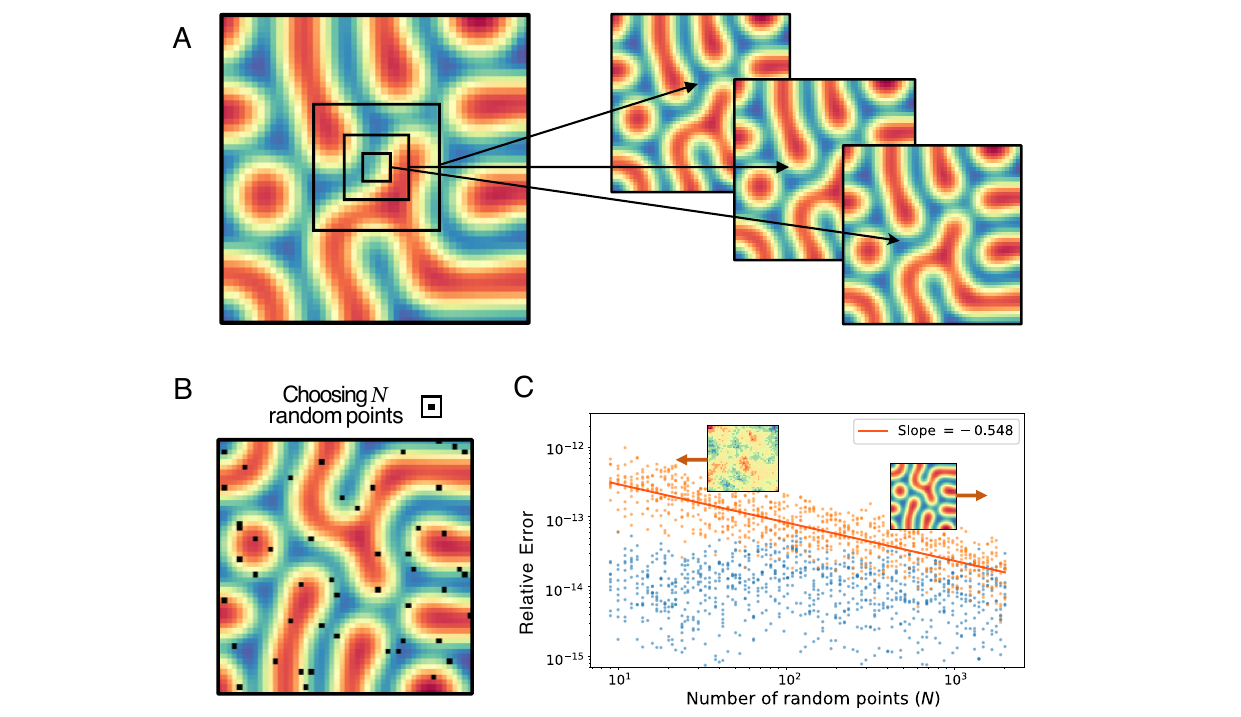}
\caption{\textbf{Effect of the number of pixels in the least squares method.} \textbf{(A)} Pattern from the FitzHugh-Nagumo model with three different cropped regions given by small squares of different size. It can be seen that a region of $3\times3$ pixels is sufficient to recover accurate enough parameters such that the predicted and original patterns (right and left respectively) are indistinguishable. \textbf{(B)} Schematic of choosing $N$ randomly selected points (black) on the Turing pattern.  \textbf{(C)} Effect of increasing the number of randomly selected pixels on the average relative error in the inferred parameters with (orange) and without (blue) added noise to the original pattern for the FitzHugh-Nagumo model. We used $N$ in the range $5$ to $2000$ ($50\times50=2500$ being the maximum possible) and sampled $10$ different sets of pixels for each $N$, and we measured the relative error for each of the inferred parameters. There is almost no effect without noise, but with noise there is a steady reduction in the relative error. Also shown is the slope of the line of best fit to the data (orange).}
\label{fig3}
\end{figure*}

To apply the LS method, we first select a model and a set of parameters, $\beta_{orig}$, that can produce a Turing pattern. Then, using the model and $\beta_{orig}$ we numerically solve the PDEs to produce a pattern similar to the ones shown in Fig. \ref{fig1}B. Once we have the patterns, we minimize $L(\boldsymbol{\beta})$. This method produces very accurate results without noise in the pattern, with an average relative error in parameters of the order of $O(10^{-15})$ or lower, which can be considered artifacts given its closeness to numerical precision. When we add noise to the pattern before applying LS, the `true' minimum shifts from $\beta_{orig}$, producing an error which is no longer a numerical artifact (Fig. \ref{fig2}A). The way we incorporate noise is by adding a matrix of normally distributed random variables with zero mean and varying standard deviation $\sigma$ to the concentration matrices $\mathbf{u}$ and $\mathbf{v}$. Since different $\mathbf{u}$ and $\mathbf{v}$ will have different ranges, the standard deviation of the noise must be relative to the range (maximum minus minimum value in the pattern) of each concentration matrix. To achieve this we employ what we called `relative noise'. If we let $R_p(\mathbf{u})$ be the range of the concentration matrix $\mathbf{u}$, so that $$R_p(\mathbf{u}) = \max_{1\leq i,j\leq N} \{u_{ij}\}- \min_{1\leq i,j\leq N} \{u_{ij}\}$$ we can define a $s\%$ relative noise to correspond to a standard deviation $\sigma = \frac{s}{100}R_p$. Once we add noise to the pattern, we can use these `noisy patterns' as input for our methods, and by changing the level of noise we can systematically investigate the difference in performance and accuracy. This will yield a different parameter set for each noisy pattern, which can be used to solve the system and obtain a new pattern. The new patterns obtained and the respective level of noise from which their parameters were inferred are shown in Fig. \ref{fig2}B. Before delving into our results, we remark that the figures only show the obtained results for the Schnakenberg and FitzHugh-Nagumo models. All the corresponding figures for the Brusselator model can be found in Supplementary Information section 3 and will be referenced when necessary.  

We considered several measures to assess the accuracy of this method. First, we looked at the average relative error in the parameters, which gives us a measure of how close the newly obtained parameters are to the original ones. A drawback of this measure is that, depending on the model, very distinct parameters can give us the same pattern, or conversely similar parameter values can yield very different patterns. Hence, a better measure for parameter accuracy would be one that quantifies how similar the pattern produced with an inferred parameter vector is to the original pattern. We found that measures like the mean squared error (MSE) are not very useful, since we do not expect the patterns to be exactly the same. This is because our main objective is to be able to recover the parameters of the model, but the final pattern is not determined solely by the parameters, but also by the initial conditions; the parameters and model determine the type of Turing pattern and its wavelength, but the position and shape are determined by the initial conditions. Hence, we do not put emphasis on obtaining the exact positions and shapes. In order to focus on the type of pattern and wavelength, we instead compare the patterns in the frequency domain. Specifically, we use the Fourier power spectrum and take the radial average to obtain a one-dimensional profile which should have a main dominant frequency for each of our patterns. This is called radially averaged power spectrum (RAPS). Example RAPS curves for different profiles are shown in Fig. \ref{fig2}C. Then, we can compute this RAPS for each pattern and use the MSE between these two profiles as a similarity measure. We will refer to this measure as RAPS difference or RAPSD for short. 

Using these two measures, we can analyze how LS performs with different levels of noise. As can be seen in Fig. \ref{fig2}B, E for both the FitzHugh-Nagumo and Schnakenberg models and in  Fig. S1A and S2A, at relative noise levels below around 0.5\% we obtain very similar patterns, even when the corresponding relative error in parameters is around 0.1\%. For larger levels of noise, both examples in Fig \ref{fig2}B begin to fail, resulting in parameters that do not produce patterns. In Fig. \ref{fig2}E we can see that the standard error in parameters (the shaded region encompassing the solid lines) behaves very differently for the two models, and its width seems to reduce at different values of relative error, so it is not caused by the scale of the error. We found that the point where this error is reduced is when the LS results are not centered around the original set of parameters anymore. To visualize this, it is helpful to note that the LS estimator is dependent on noise. Hence, different realizations of the corrupted pattern at the same level of noise will produce different LS estimators, which will produce different patterns, as can be seen with $1\%$ noise in Fig. \ref{fig2}B. We can either obtain a pattern or a constant solution, showing how the obtained parameters are at the boundary of the Turing region. These different estimators can be thought of as a set of points in parameter space. Subsequently, the reduction in standard error occurs as the set of points moves further away from the original set of parameters, as described in the sketch in Fig. \ref{fig2}E). Note that since we are using LS, instead of computing the relative error in $\hat{\beta}$, it could be argued that we can use the statistical properties of the LS estimator and obtain an expression for its variance. Indeed, the relative error we are computing is proportional to the standard deviation of $\hat{\beta}$. Hence, if we had an expression for the variance, then the errors would be easily computed. However, this is not an easy task since the usual LS assumptions do not hold in our case (noise is not additive and does not have a constant variance, and the design matrix is noisy). Applying the statistical properties of the LS estimator results in errors that do not match our data, so we decided to numerically estimate the relative error. 

In Figs. \ref{fig2}D and S2B we notice discrete jumps when comparing RAPS profiles, which we can match to behaviors in the patterns. By comparing with different noise levels in Fig. \ref{fig2}B, we found that the last jump separates inferred parameters that produce a pattern and inferred parameters that do not. The other jumps are less clear and have to do with the patterns changing wavelength and scale. For example, for the Schnakenberg model we can notice a slow linear trend, which shows that the range of the patterns is slowly changing. This is followed by a jump around $0.4\%$, which is the point where the spots have grown too much and the pattern changes to fewer dots (see difference in spot number between $0.3\%$ and $0.6\%$ noise). For the FitzHugh Nagumo model, we notice more jumps and it is less clear what these represent. We saw that the general behavior is conserved for different patterns with the same parameters, but the position of jumps seem to be dependent on the initial conditions. Note that the main advantage of this measure is that it allows us to see when changes in the pattern occur. Nevertheless, this measure is computationally more expensive than computing the relative error in parameters, since we need to numerically solve for the pattern. Hence, we still use the relative error in parameters.

Our results show that the LS estimators are not robust to the levels of noise we aim for, but in the absence of noise the method works well with little data. As can be seen in Fig. \ref{fig3}A, even if we crop the original pattern and use a $4\times4$ pixel region, we still obtain the same patterns. A natural question is to find the minimum number of pixels. Given the simplicity of the method, to answer this question we need only consider the formulation of LS. A full explanation is provided in the \nameref{sec:methods} section but we give a brief explanation here: given the design matrix $X$, which is determined by the model, we need $X^TX$ to be invertible. This condition will be satisfied when $X$ has enough rows (a pixel corresponds to two rows) so that it is a full rank matrix, making the minimum amount of pixels be model dependent. For the Schnakenberg and Brusselator models, 2 pixels are enough, whereas 3 pixels were required for the FitzHugh-Nagumo model.

Next, we investigated how much the addition of new pixels changes the relative error in the parameters. To do this, we sampled $N$ pixels from the pattern using a two dimensional uniform distribution (Fig. \ref{fig3}B), which we used to infer a new set of parameters. In Fig. \ref{fig3}C we show a scatter plot of the mean relative errors for each $N.$ We observed that the effect of the number of pixels on the relative error is different depending on whether the original pattern was corrupted with noise or not. Without noise (blue) we hardly see any improvement, but with noise (orange) there is a more drastic improvement in the parameters, which is expected as the noise has less effect the larger the sample. As the amount of randomly selected pixels is increased, the relative error approaches that of no added noise with a power law, as can be seen in the log-log plot in the upper right corner of Fig. \ref{fig3}C. The exponent of this power law is close to $-\frac{1}{2}$, which agrees with previous results on the LS convergence with $N$ \cite{exponentconvergence}. For an intuitive mathematical explanation of this convergence see Supplementary Information Section 4. We remark that the fluctuations in relative error without noise (or for large $N$ with noise) are numerical instabilities, since they are also observed if we use all the data and merely change the order of the input. Similarly to this, we also investigated this effect in the RAPSD for the Schnakenberg model in Fig. S3. We saw that the observed behavior was dependent on the initial level of noise chosen: for a noise around $0.03\%$, we observe a similar power law as with the relative error, with a slope again close to $-\frac{1}{2}$. For a noise of $0.25\%$, we observed the same jumps as can be seen in Fig. \ref{fig2}D, with a convergence towards the same value shown in the figure for this level of noise. (For more discussion see Supplementary Information Section 5.)

We also attempted to apply LS to a design matrix defined by a model not used to generate the pattern. For example, we produce a pattern with the Brusselator model but we define the LS minimization using the Schnakenberg model. This is what we called `mixing' the models, and the goal is to check if the models are flexible enough to produce the same pattern from one another. We found that this did not work even between models capable of producing the same type of pattern, and instead produced parameters that gave no Turing patterns at all (see the Discussion section). We can see that this result enforces the idea that a model is like a key, and when we have the key, the pattern is simply a barcode which is easy to read and gives us the model parameters. In summary, the LS method works well without noise in the the data, as it returns the exact parameters (with some numerical error), even when we have very little data (2-3 pixels depending on the model).

\begin{figure*}[t]
\includegraphics[width=\textwidth]{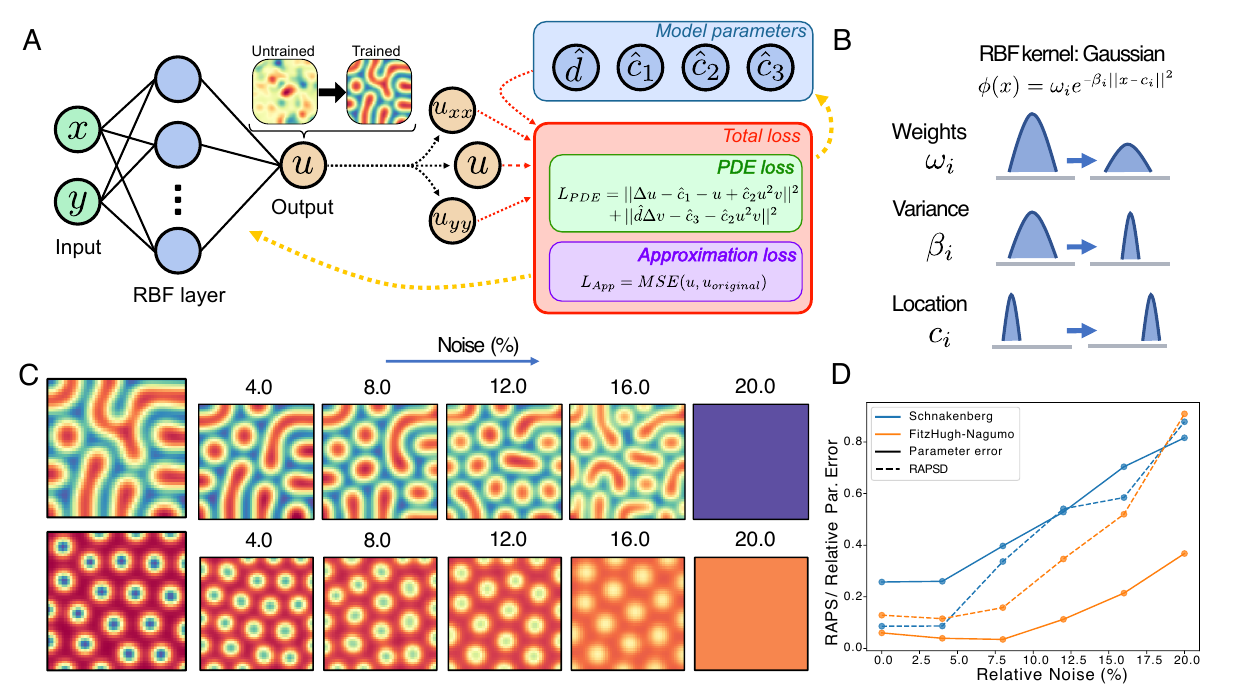}
\caption{\textbf{Physics-informed neural networks for parameter inference.} \textbf{(A)} Architecture of RBF-PINNs, where the input is space coordinates ($x,y$) and the output is the pattern. Input is shown in green, variables which are trained in blue and input to the losses in yellow. Red arrows show denote usage in the loss and yellow arrows backpropagation. From the network the partial derivatives can be efficiently computed using automatic differentiation and used in the PDE loss, where the PDE parameters are also network parameters. \textbf{(B)} Illustration of the three parameters of the Gaussian kernel and their interpretation. \textbf{(C)} Results from RBF-PINNs with different levels of added noise. After adding noise to the Turing pattern, the network is used to obtain a parameter set, which is subsequently used to predict the pattern. Patterns to the left are the original ones. \textbf{(D)} Relative error in the parameters and the RAPS difference for the parameters and patterns used for (C).}
\label{fig4}
\end{figure*}
\subsection*{Physics-informed neural networks} 
Physics-informed neural networks (PINNs) are a neural network architecture in which the network serves as a function approximator with two loss functions: the first compares the output of the network to the numerical simulation of the PDE (the pattern in our case) and makes sure the network outputs the correct values; the second tries to make the output of the network a solution of the PDE (the physical law) by both optimizing the network and optimizing the parameters of the PDE (the estimation)\cite{KarniadakisPINNs}. Using these two losses, the network approximates the solution as well as learns the optimal set of parameters for which the Turing pattern is a steady-state solution. This approach is computationally very efficient, since it only adds a few parameters to the network which usually trains thousands. Hence, the extra computational cost is minimal, leading to excellent scaling with the number of parameters in the PDE. A peculiarity of this method is that if we have no noise in our input patterns, we can let the approximation overfit the data. This is because the approximation will become better and this will improve the parameter optimization as well. 
 
As function approximation we used a radial basis function neural network (RBFNN), since it is especially suitable for data with regularities such as evenly spaced peaks and valleys like our Turing patterns. This network only has one hidden layer (aside from input and output) defining the kernel. We can think of each kernel as adding a (unnormalized) distribution (Gaussian in our case) at a given location with a given variance and weight. These three are the only types of parameters of this network. Hence, the amount of nodes is the amount of kernels that we have, and during training these will change their location and variance, shrinking or growing to approximate the pattern. A representation of the three parameters and their effects are shown in Fig. \ref{fig4}B.

As with our previous method, without added noise the network can approximate the pattern up to an arbitrary accuracy \cite{RBFunivApp}. This means that we can reach the same levels of accuracy as with the other method, but with a caveat: we need to train the network for a long time to reach this level of accuracy and we would need a large number of nodes in the hidden layer. This is the reason behind the large error in Fig. \ref{fig4}D without added noise, since we use the same amount of nodes for all noise levels. Note the performance  can easily be improved without noise, but as our main focus is robustness to noise, we did not tune any parameters to any specific level of noise, except for the number of nodes and the variance of the kernels, and we use exactly the same network. Once we apply noise, our results with this technique show an improved robustness when compared to LS, as can be seen in Fig. \ref{fig4}C. For both the FitzHugh-Nagumo (top) and Schnakenberg (bottom), we obtain very similar patterns up to a noise level of 12-16\%. For the Brusselator model, shown in Fig. S1B, we can see similar results. Comparison with Fig. \ref{fig4}D and Fig. S2C and D shows that the relative error in parameters is higher in the Schnakenberg than in the FitzHugh-Nagumo model, but this error is much lower in the Brusselator model, indicating that this model is more robust to noise. Nevertheless, from Fig. \ref{fig4}C we can see that there is not a direct relationship between relative error and similarity between the patterns (RAPSD), since for low noise levels (0-4\%) the relative error in the Schnakenberg model is higher and the RAPSD is lower than the FitzHugh-Nagumo model, but for higher levels the RAPSD of the Schnakenberg model becomes higher. For convergence plots of some of the model parameters and losses of the network see Fig. S4 in Supplementary Information Section 6.

By looking at the RAPS profiles we find the same discrete jumps that we saw with the LS method. In particular, a well pronounced final jump is observed, which we argued before indicates the transition from pattern to no pattern. By comparing Figs. \ref{fig2}D and \ref{fig4}D, RBF-PINNs with 0-4\% noise results in RAPS values corresponding to around 0.3-0.5\% for LS while the points with 8-12\% noise correspond to the ones around 0.6-0.9\%. This means that RBF-PINNs can infer parameters and predict patterns with around 10-20 times more noise than LS for the Schnakenberg and FitzHugh-Nagumo models. In the case of the Brusselator this number is even higher, as can be seen by comparing Fig. S2B and D, with RBF-PINNs being able to predict patterns with up to 40 times higher noise levels.

\subsection*{Application to data from chemical patterns}

Having demonstrated how inference by PINNs shows strong robustness to noise, the next step is to apply this method to experimental data. We chose chemical Turing patterns as an example because patterns are more stable and robust than biological patterns, and corresponding models are well established. Specifically, we consider here the chlorine dioxide–iodine–malonic acid (CDIMA) system used to study the impact of 2D growth on patterns formation \cite{ChemPat2019}. This reaction shows a photosensitivity that was utilized to produce a time series of snapshots of the pattern at different times in a radially growing domain using a mask. Since our method focuses on the steady state pattern, we discarded all the time points except for the last one, focusing on the stable central region away from the boundary.  

The experiments were modeled using a Lengyel-Epstein two-variable model \cite{Epstein_Chemical_Model}, modified to incorporate the effects of illumination \cite{Illumination_pattern}. Since we are only interested in the pattern, which is shown in the dark region, we will use the original model without illumination. We found that this model, although already non-dimensional, was problematic for parameter inference. This was because in one of the equations all terms have a parameter, which can lead to a trivial solution where all parameters are zero. To solve this, we derived a new non-dimensionalization:

\begin{equation}
    u_t = d \Delta u +c_1- u - c_2\frac{4 u v}{1+u^2}   \qquad v_t = \Delta v + u - c_2\frac{4 u v}{1+u^2},\label{NondimChem}\\
\end{equation}

which we used to fit the data with our RBF-PINN. Before stating our results, there are a few complications worth mentioning. First and foremost, the scale of the patterns is unknown, since the original data is an image with pixel values ranging from 23 to 255. Secondly, we have two concentrations on the model ($u$ and $v$) exhibiting a pattern, but only one experimental pattern. In order to solve both these problems, we define a free-scale variable $W$, which is a rescaled version of the original pattern in the range $[0,1]$. In order to obtain the model concentrations, we assume that there is a linear map from $W$ to $u$ and $v$, which we can write as:

\begin{equation}
    u = W \kappa_u + \gamma_u \qquad v = W \kappa_v + \gamma_v,\label{Scaling}\\
\end{equation}

where $\kappa_x$ and $\gamma_x$ for $x =u,v$ are scale and shift parameters respectively. The assumption of the existence of this linear map can be justified by the fact that patterns usually are either in phase (positive $\kappa$) or out of phase (negative $\kappa$).

\begin{figure*}[t]
\includegraphics[width=\textwidth]{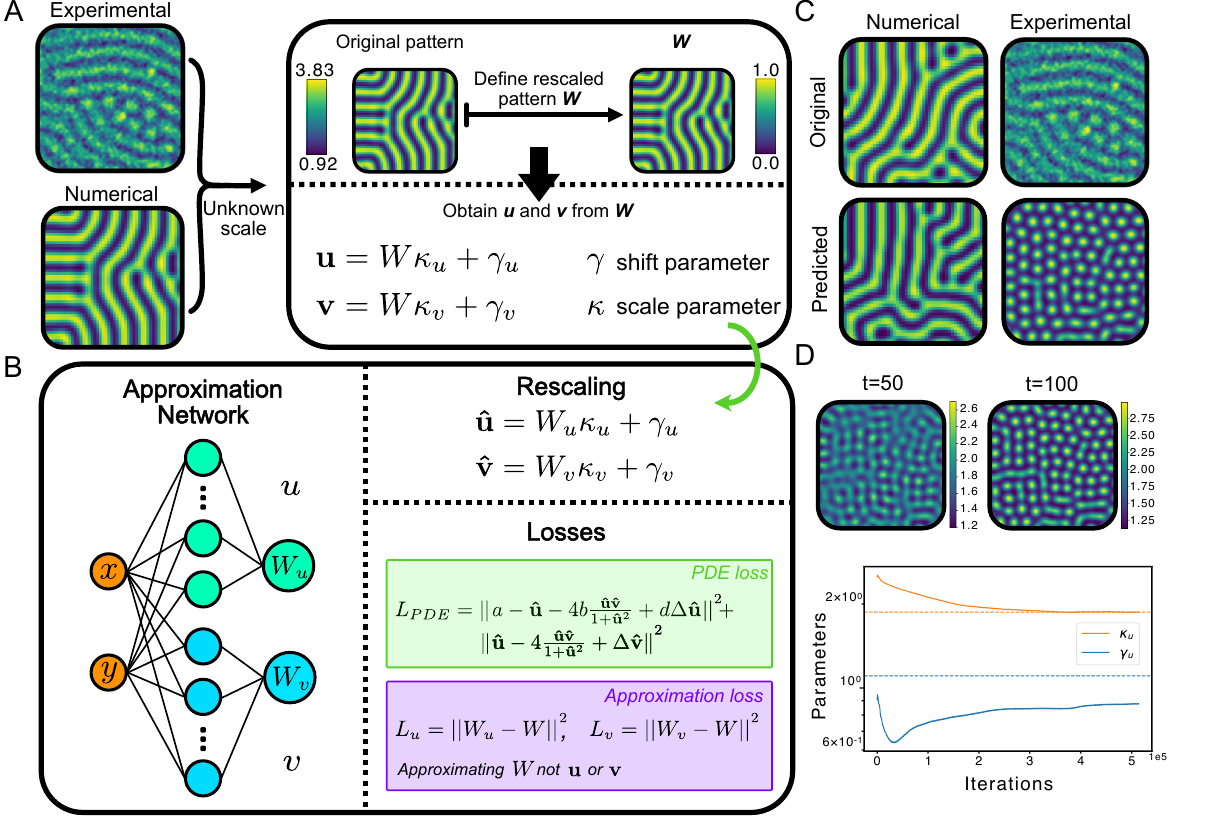}
\caption{\textbf{Application to chemical patterns.} \textbf{(A)} Explanation of scaling procedure, showing numerical and experimental patterns and how the scaling to the free-scale variable and the rescaling using the shift and scale parameters are performed.  \textbf{(B)} Architecture of RBF-PINNs for the experimental case, with the division into the $u$ and $v$ approximation, the rescaling and the different losses. \textbf{(C)} Results from RBF-PINNs to the numerical pattern (first column) and the experimental pattern (second column). The top images show the original patterns, while the bottom images show the predicted patterns using the inferred parameters and our numerical solver. \textbf{(D)} Time evolution of simulated experimental pattern showing how labyrinths are present at the initial time points and plot of convergence of scaling parameters.}
\label{Fig5}
\end{figure*}

As before, our network will consist of two independent subnetworks which will approximate $W$ individually. From these two approximations we will recover $u$ and $v$ using the scaling in Eq.\ref{Scaling}, so we will call these approximations $W_u$ and $W_v$, respectively. The reason for using two approximations for the same variable is that usually patterns are not perfectly in phase or out of phase. Hence, by separating these into two variables we allow for flexibility in the method to make adjustments to each pattern individually. Another difference with the network used previously is that the scaling parameters are added as part of our PDE loss, and hence are optimized with the rest of parameters, as portrayed in Fig. \ref{Fig5}B.

We firstly tried our inference method using numerical `data' from a simulation of the pattern. This gave us two patterns, but we only used one of them to make this more similar to the experimental case. As a proof of concept, we started by fixing the scaling parameters to the best combination possible for both $u$ and $v$, and obtaining the rest. The predicted pattern is shown in the first column of Fig. \ref{Fig5}C (bottom), together with the original one (top), and we can see a good correspondence between them. Specifically, both show labyrinths on a similar scale. We obtained a RAPS of 0.2941 for the two numerical patterns, which is close to the value obtained for the FitzHugh-Nagumo model in Fig. \ref{fig4}D below 16\% of noise.

Encouraged by the numerical result, we tried our approach on the experimental pattern. We initialized all parameters close to the values from the numerical pattern and also used the same spatial dimensions. The results are shown in the second column of Fig. \ref{Fig5}C. We can see that the predicted pattern (bottom) is not particularly close to the experimentally observed one (top). The obtained pattern also seems to show spots instead of labyrinths, like the experimental one. Nevertheless, when inspecting the time evolution of this pattern in Fig. \ref{Fig5}D, we can observe that at the beginning the spots are connected and later on they move apart. By looking at the experimental videos, we can actually see this happening too - at the beginning we can only observe labyrinths but later on spots begin to appear. This seems to point to the fact that the experimental pattern has not completely converged and that if left to evolve for longer, it would end up in a steady state with only spots. Furthermore, in Fig. \ref{Fig5}D (bottom) we see the convergence of the scaling parameters for $u$ (solid lines) along with the values from the original numerical pattern (dotted lines). We can notice that they match quite well, with a slight mismatch of the shift parameter which is probably due to feedback in the training with other parameters. We could also infer that the chemical pattern that was produced experimentally was most likely an activator, $I^-$, which we later confirmed \cite{Turing_pattern_activator}. This is because depending on whether the initial concentration belongs to $u$ or $v$ in the model, one non-dimensionalization will work better than the other; in our case only $u$ worked. The network training is further discussed in the Supplementary Information section 7.

In summary, our method can be applied to experimental data with only minor manual intervention. Furthermore, we were able to obtain insight into the chemical pattern e.g. its closeness to a spotty steady state, due to the parameters we obtained from the method. Our method predicted the correct identification of the pattern with the activator chemical species.

\subsection*{Discussion} \label{sec:discussion}
We investigated two methods to solve the inverse problem in Turing patterns. The first was based on the least squares (LS) method, and was found to work very accurately, with relative errors under $10^{-15}$, when no noise is added to the patterns. We also proved that it works even with very small quantities of data: 2-3 pixels from the discrete patterns were enough, and that this depends on the model equations. We also proposed another method based on neural networks, that we called RBF-PINNs since it is a mixture of PINNs and RBF neural networks. Without noise, this method can be compared with LS, since the universal approximation theorem assures that we are able to approximate any function up to any given accuracy \cite{RBFunivApp}. The main advantage of this method comes when we consider a noisy pattern, since in this case our method was capable of recovering similar parameters to the original ones with up to 12-16\% relative noise.This allowed us to use this method for experimental chemical patterns, and obtain a set of parameters that gave us new insight about the possible future behavior of the patterns obtained experimentally. 

Each of these methods has its strengths and weaknesses. The main disadvantage of the LS method is its sensitivity to noise, which is usually present in most biological systems and data. This was remedied by our RBF-PINN method at the expense of a computational cost several orders of magnitudes larger than LS, taking nearly an hour to train on a HP Z8 G4 Workstation. Interestingly, the inferred parameters without noise are not as accurate as LS, which can be observed by comparing the relative errors in Figs. \ref{fig2}D and \ref{fig4}D. This performance can be further optimized by using more nodes in the network. Also the robustness to noise can be improved. Both our approaches used a square grid of $50 \times 50$ pixels, but as we increase the amount of pixels, the performance of both methods in presence of noise becomes more robust. We can also run the optimization of the loss functions for longer to obtain a better convergence, or we can alter the architecture to make it more specific to our patterns. In this paper, we use exactly the same network architecture for all patterns, only changing the variance of the starting RBF kernels since the patterns have a different wavelength and scale, and the number of nodes. However, we could get more accurate results if we were to increase the number of nodes in the network, although we would have to be careful about overfitting to noisy data. 

When comparing with other attempts at solving the inverse problem for Turing patterns, we should first make the distinction between approaches that treated noise and those which did not. For the approaches that did not consider noise \cite{Kazarnikov2020}, the simple LS performs better than this Bayesian-based approach since it does not need any training and it is more generalizable, since we only need knowledge of one pattern and not a library of patterns. Compared to the ones that did consider noise \cite{Garvie2010,CampilloBayes}, the main difference is that we do not require knowledge of the initial conditions or of transient data (dynamics). These are assumptions that would not hold for biological data, which usually is scarce and with high levels of noise, so our approach is to the best of our knowledge the first one that tackles all of these issues (small data and noise) at the same time. Recently, the inverse problem was addressed with neural networks but with some limitations: in \cite{TuringPinns} PINNs were applied to Turing patterns but without considering noise and in \cite{Rao2023} an architecture based on recurrent neural networks was developed but without testing it on actual data.

Despite the successes of our methods to deal with small data and noise, there are some limitations worth mentioning. First, we did not use domain growth for our work, which would be more biologically relevant. This could be implemented by considering more traditional PINNs that take time into account. We did not focus on time series data, although our LS approach easily extends to this (see Supplementary Information section 9), and the more traditional PINN structures are very adequate for data changing in time. Our approach to incorporate the scaling in the chemical patterns, although effective for this case, was found to be very sensitive, since parameters could interfere with other parameters during training. A more sophisticated encoding of the scaling used could potentially provide a better distinction in the parameters and yield better and more robust results. Looking at possible biological applications, a problem would be the simplicity of the models, which are just cartoon representations of the actual biological networks as encountered e.g. in developmental biology. To inform biological experiments, future models need to be complex enough to have meaningful connections to experimental observables, while being simple enough to avoid the ‘curse of dimensionality’. At the same time, experiments should show a good connection with the simulations in the models, as otherwise the approaches shown in this work would not be useful. As a remark, we only focused on finding the parameters, but it would also be interesting to extend this to model selection. Existing methods include graph networks \cite{SymbRegPhys}, embedding theory and diffusion maps \cite{PDESiettos,ChemotaxisSiettos}.

Our approaches open up new ways of connecting mathematical models to experimental patterns. In particular, RBF-PINNs worked for noise levels comparable to biological data, and proved to be useful at elucidating properties of chemical patterns. Hence, this method could potentially be applied to infer parameters for biological candidate models, “proving” that a model is capable of reproducing observed patterns. This type of model selection could ultimately aid the rational design of synthetic tissues with patterns for downstream templating and added functionality \cite{Davies2016, Toda2020}. In conclusion, we hope our machine learning approaches to solving inverse problems stimulate new research into unraveling pattern formation in biology, chemistry, and bioengineering.

\section*{Methods}\label{sec:methods}
All algorithms and methods were run in Python on a HP Z8 G4 workstation with a Intel® Xeon(R) Gold 6128 CPU @ 3.40GHz × 24  processor and a Quadro RTX 6000/PCIe/SSE2 GPU. Packages and requirements of the environments are explained in the Supplementary Information section 8.
\subsection*{Numerical simulations}

All numerical simulations of the models were performed on a $50\times 50$ discrete grid by applying a center difference approximation to the second-order derivatives in the Laplacian, hence transforming the PDE Eq. \ref{general_model} into an ODE model:

\begin{align}\label{ODE_model}
\begin{split}
    \frac{du}{dt}=& D_u \frac{u_{i,j-1}+u_{i,j+1}+u_{i-1,j}+u_{i+1,j}-4u_{i,j}}{(\Delta x)^2} + f(u,v) \\
    \frac{dv}{dt}=& D_v \frac{v_{i,j-1}+v_{i,j+1}+v_{i-1,j}+v_{i+1,j}-4v_{i,j}}{(\Delta x)^2} + g(u,v) \\
\end{split}
\end{align}

Here, we assumed that the step in the $x$ direction is the same as the step in the $y$ direction, i.e $\Delta x= \Delta y$. With this approximation, we use a numerical solver to forward integrate these equations for a sufficiently large amount of time that ensures that the system has converged to a steady state, which is the Turing pattern. 

\subsection*{Least squares}

As described in the section on LS, in order to write down the solution we first need to write our problem in matrix form. We will do this here with the Schnakenberg model as an example. Using the same notation as before, we let $u_{ij}$ and $v_{ij}$ be our discretized Turing patterns in matrix form, for $i,j=1,2..., N$, with $N$ being the number of columns and rows (as we have a square grid). Writing Eq. \ref{NonDimS} using this notation and at steady state we obtain:

\begin{equation}
    u_{ij}-\Delta u_{ij} =  c_1  +c_2 u_{ij}^2 v_{ij}  \qquad 0  = d \Delta v_{ij} + c_3 - c_2 u_{ij}^2 v,\label{NonDimSDiscrete}\\
\end{equation}

where we collected the terms not multiplied by a parameter to the left. This is a total of $2\times N^2$ equations, which we can write in matrix form as:

\[
\begin{pmatrix}
    u_{11}-\Delta u_{11}\\
    u_{12}-\Delta u_{12}\\
    \vdots\\
    u_{NN}-\Delta u_{NN}\\
    0\\
    0\\
    \vdots\\
    0
    
    \end{pmatrix}=
   \begin{bmatrix}
     1 & u_{11}^2 v_{11} & 0 & 0 \\
     1 & u_{12}^2 v_{12} & 0 & 0 \\ 
     \vdots & \vdots & \vdots & \vdots \\   
     1 & u_{NN}^2 v_{NN} & 0 & 0 \\
     0 & -u_{11}^2 v_{11} & 1 & \Delta v_{11} \\
     0 & -u_{12}^2 v_{12} & 1 & \Delta v_{12} \\
     \vdots & \vdots & \vdots & \vdots \\   
     0 & -u_{NN}^2 v_{NN} & 1 & \Delta v_{NN}
     
   \end{bmatrix}
   \begin{pmatrix}c_1 \\c_2 \\ c_3 \\ d \end{pmatrix} \]

We will write this equation as $y = X\beta$, and will call $y$ the vector of independent variable and $X$ the matrix of dependent variables. It is also customary when performing regression to call $y$ the vector of outputs and $X$ the design matrix. This is the starting form of the LS formulation \cite{hastie01statisticallearning}. By defining the error vector and finding its minimum, the solution for the optimal parameters can be written as:

\begin{equation*}
    \beta = (X^TX)^{-1}X^Ty
\end{equation*}

Note that for this approach to work we need $X^TX$ to be invertible, which is equivalent to $X$ having full rank. Based on this necessary condition, we can obtain the minimum amount of pixels needed for the solution to be well defined. In the case of the Schnakenberg model, we can see that two pixels are enough, since it gives us a $4\times 4$ matrix in the form:

\[M=
   \begin{bmatrix}
     1 & a & 0 & 0 \\
     1 & b & 0 & 0 \\ 
     0 & -a & 1 & c \\
     0 & -b & 1 & d   
   \end{bmatrix}
 \]

which for most $a,b,c,d$ will be full rank. There are cases (e.g. $c=d$ or $a=b$) where $M$ will not be full rank, but for randomly selected pixels in the pattern these cases are unlikely. Likewise, we can easily see that this condition gives two pixels for the Brusselator model (since we have two independent $2\times 2$ submatrices) and three for the FitzHugh-Nagumo model (since we have a submatrix of $3\times 3$).
\subsection*{Neural networks}
For the neural network, as described in the \nameref{sec:results} section, we combined PINNs (i.e. using the PDE equations in order to improve the approximation and having the parameters as trainable variables in the network) with RBF neural networks to obtain what we called RBF-PINNs. The architecture is very simple since the network is made of only three layers. The input of the network is a vector representing a spatial location, in our case $(x,y)$, since we only have two dimensions, and the output is $u(x,y)$. We have one network for each $u$ and $v$, to allow for more flexibility in the model, but we train them together. Between the input and the output layer we have a single hidden layer, with an activation layer representing the RBF kernels, and the input being fed directly to this activation layer by having all weights set to 1, which means that the amount of trainable parameters is in the order of $3N$ if we have $N$ nodes. These RBF kernels can be written as:

\begin{align}\label{RBF_kernel}
\begin{split}
    \phi_i(\mathbf{x}) =  w_i e^{-\beta_i ||\mathbf{x}-\mathbf{c_i}||^2} \quad \text{for} \quad i=1, \:2,...\: M 
\end{split}
\end{align}

where $M$ is the total number of nodes in the network. As explained previously and portrayed in Fig. \ref{fig4}, $\beta_i$ works as the variance, $c_i$ is the location of the $i^{th}$ kernel and $w_i$ its weight or importance. Each node in the network represents a different kernel, so the only way to change the architecture of the network to improve the accuracy of the approximation is to increase the amount of kernels by increasing the amount of nodes. The kernel parameters were set randomly, $\mathbf{c}$ was set to a 2D uniform random variable in the spatial range of the pattern, and the network weights were initialized using Glorot initialization. The variance was initialized depending on the pattern, since the variance of the Gaussian kernels has to be on a similar scale as the pattern wavelength.

Our network has two main losses. One is an approximation loss, which compares the output of the network to the pattern. If we consider our network as a function, we can write it as $\Phi^u(\textbf{X})$ where $\mathbf{X}$ is the space dimensions where we have the pattern values $\mathbf{u}$. We will from now on just write this as $\Phi^u$ and similarly for $\mathbf{v}$. Subsequently, we can write the approximation loss for $\mathbf{u}$ as:

\begin{align}
\begin{split}
    L_{app,u} = \frac{1}{N^2}\sum_{i,j} ||\Phi^u_{i,j}-\mathbf{u}_{i,j}||^2
\end{split}
\end{align}

Note that this is equivalent to computing the mean squared error (MSE) between the data and the approximation. We mentioned the RAPS loss before, and it could be used here for the network training as well, but since we care about the pattern being accurate, the MSE works fine. However this could be a possible improvement since it should cancel the noise to some extent.

The second loss is the PDE loss, which enforces that the output has a small PDE residual and at the same time improves the parameters so that the data gives a better fit to the equation:

\begin{align}
\begin{split}
    L_{PDE} = \frac{1}{N^2}\sum_{i,j} ||\beta_1 \Delta \Phi^u_{i,j} +f(\Phi^u_{i,j},\Phi^v_{i,j},\boldsymbol{\beta})||^2+||\Delta \Phi^v_{i,j} +g(\Phi^u_{i,j},\Phi^v_{i,j},\boldsymbol{\beta})||^2
\end{split}
\end{align}

We also have a loss for the diffusion term, to make sure that the second derivatives of the approximation are accurate. The final structure used for the results in the paper has 60-120 nodes depending on the model. For training, we use batch training with batches of 128 elements for 200,000 iterations and Adam optimizer, taking a total of less than an hour to run. The first 10,000-20,000 iterations are used to approximate the function, since we only update the weights using $L_{app}$, and after this we change to minimizing both losses, taking care of using weights before each loss so that the approximation does not get completely changed by the addition of the new losses. When training the approximation, we use information on the whole pattern, but when we switch to the $L_{PDE}$, we only use interior points so that inaccuracies in the diffusion do not present a problem. 

The structure of our network is relatively simple and the RBF network guarantees a smooth output. However, to avoid overfitting we can use a small number of nodes or kernels. This is because overfitting is only a problem for the approximation part of our network (that is where our data is used). Furthermore, our patterns have a larger wavelength than the noise. By selecting a variance or amplitude parameter to be of the same order as our pattern, we keep the network from incorporating most of the noise. On top of that, we redefine the weights of the network every 2,000 iterations so that all of them are on a similar scale. This is useful since it is possible that some weights converge faster than other, and so would eventually have a very small loss if the original weights were used. 

\section*{Acknowledgments}
We thank Roozbeh H. Pazuki for invaluable support and technical comments, Martina Oliver Huidobro for stimulating discussions and Milos Dolnik for explanations of his experiments. This research was funded through a studentship from the Department of Life Sciences at Imperial College London.

\section*{Declaration of interests}
The authors declare no competing interests.

\section*{Author contributions}
 A.M.G. and R.G.E. designed, and A.M.G. implemented and performed the theoretical computational approach and data analysis. Both authors wrote the paper.
 
\bibliographystyle{vancouver}
\bibliography{./refs.bib}
\end{document}